\documentclass[aps,twocolumn,prb,amsmath,amssymb,showpacs,notitlepage,superscriptaddress]{revtex4-1}
\usepackage{graphicx,bm}

\begin{document}
\title{Detection of entanglement by helical Luttinger liquids}

\author{Koji Sato}
\affiliation{Department of Physics and Astronomy, University of California, Los Angeles, California 90095, USA}
\affiliation{WPI Advanced Institute for Materials Research, Tohoku University, Sendai 980-8577, Japan}
\author{Yaroslav Tserkovnyak}
\affiliation{Department of Physics and Astronomy, University of California, Los Angeles, California 90095, USA}

\begin{abstract}
A Cooper-pair or electron-hole splitter is a device capable of spatially separating entangled fermionic quasiparticles into mesoscopic solid-state systems such as quantum dots or quantum wires. We theoretically study such a splitter based on a pair of helical Luttinger liquids, which arise naturally at the edges of a quantum spin Hall insulator. Equipping each helical liquid with a beam splitter, current-current cross correlations can be used to construct a Bell inequality whose violation would indicate nonlocal orbital entanglement of the injected electrons and/or holes. Due to  Luttinger-liquid correlations, however, the entanglement is exponentially suppressed at finite temperatures.
\end{abstract}

\maketitle

\section{Introduction}
Controlled generation, manipulation, and detection of entangled quantum states are crucial ingredients for quantum computation,\cite{nielsenBook00} quantum teleportation,\cite{bennettPRL93} and quantum cryptography.\cite{bennettIEEE84,*ekertPRL91,*gisinRMP02} The Einstein-Podolsky-Rosen (EPR) thought experiment similarly relied on the control of entangled states.\cite{einsteinPR35} One of the ways to test quantum entanglement is to observe a violation of a Bell inequality.\cite{bellRMP66,*clauserPRL69} Although this has been achieved with high accuracy using entangled-photon sources,\cite{aspectPRL81,*aspectPRL82,*fransonPRL89,*kwaitPRL95} performing such experiment with electrons is a challenging task, because of electron-electron interactions and dephasing due to the solid-state environment. Nonetheless, Bell tests based on electron spin entanglement,\cite{kawabataJPSJ01,*chtchelkatchevPRB02,*lebedev-lesovikPRB05} orbital entanglement,\cite{samuelssonPRL03} and electron-hole entanglement\cite{beenakkerPRL03,satoPRB14} have been proposed, where a Bell inequality is built upon charge current correlations.

Sources of entangled particles and mechanisms to spatially separate them are essential requirements for performing a Bell test. This task can be achieved by a Cooper-pair (CP) splitter, which can spatially separate a spin-entangled CP by sending a weak current from a superconductor (SC) into a pair of quantum dots, wires, or carbon nanotubes.\cite{lesovikEPJB01,*recherPRB01,*recherPRB02,*benaPRL02,*recherPRL03,*samuelssonPRB04} An $s$-wave SC provides an excellent source of spin-entangled electrons from CPs, which are condensed at the Fermi level of its ground state. Spatial separation of a CP can be achieved through crossed Andreev reflection,\cite{beckmannPRL04,*russoPRL05,*weiNatPhys10} as has been recently demonstrated using double quantum dot structures in single-wall carbon nanotubes\cite{herrmannPRL10,*herrmannCM12} and InAs semiconductor nanowires.\cite{hofstetterNature09,*hofstetterPRL11} The efficiency of CP splitting was shown to approach unity,\cite{schindelePRL12} which encourages further pursuit of superconducting heterostructures toward Bell tests and, in time, scalable quantum measurements.

\begin{figure}[pt]
\includegraphics[width=0.8\linewidth]{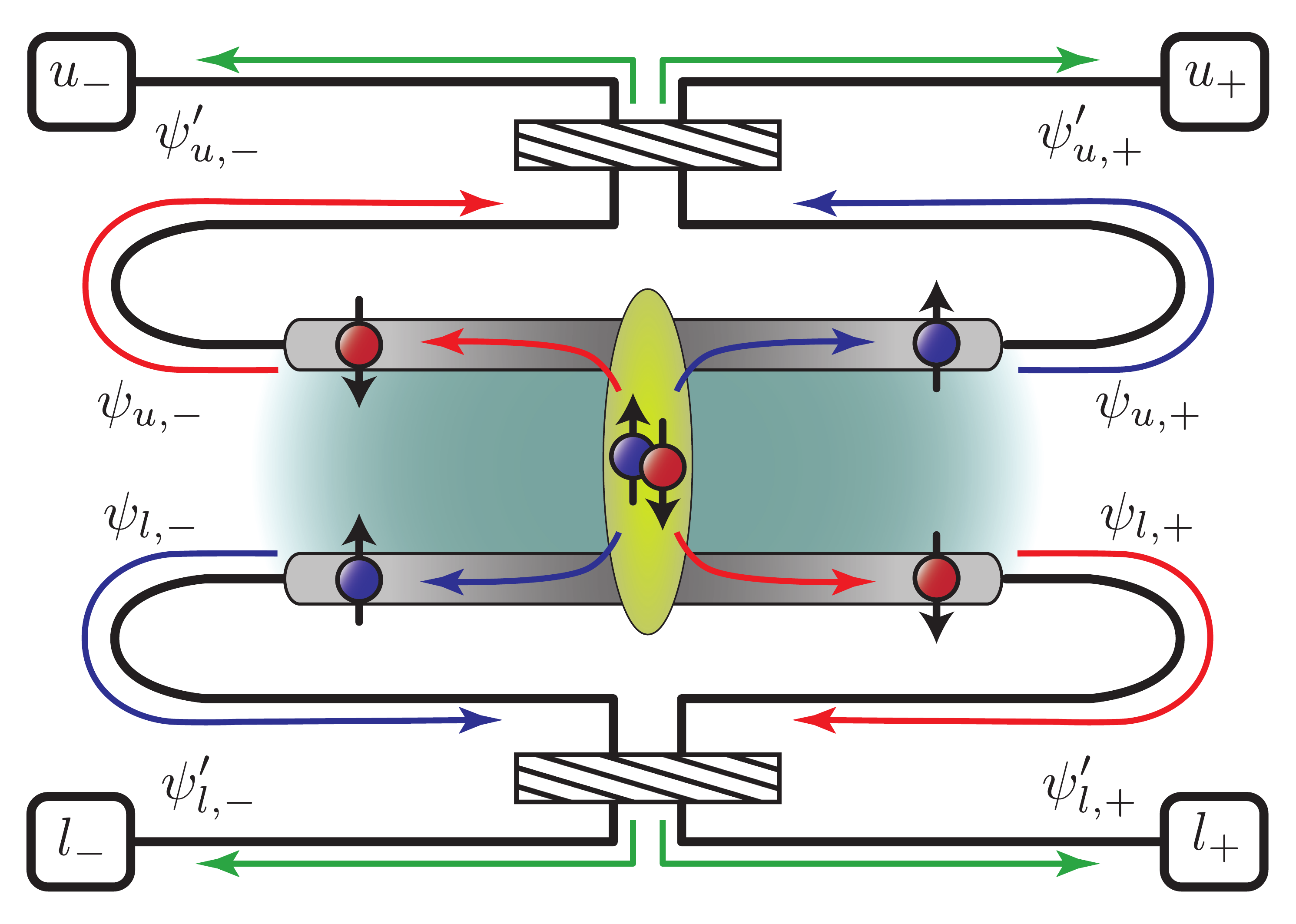}
\caption{An $s$-wave SC is coupled to a QSHI. Two electrons forming a CP split into top and bottom helical edge states. The electron-electron interaction is finite in the grey regions around SC and vanishes outside of these regions. Two beam splitters are formed at the edges, which are indicated by the striped regions. The charge currents are detected at the end points labeled by $u_\pm$ and $l_\pm$. $\psi_{n,r}$ with $n=u,l$ and $r=\pm$ indicate the incoming electron states moving to the right (+) and left (-) along the upper ($u$) and lower ($l$) edges, and $\psi_{n,r}'$ are the outgoing states perturbed by the beam splitters.}
\label{fig1}
\end{figure}

After a spin-entangled pair of electrons is spatially separated, their spins need to be read out. Traditionally, the information on spin is extracted by a spin-to-charge conversion,\cite{elzermanNature04,hansonRMP07} where a spin state is directly related to charge current via spin filtering controlled by a local magnetic field or exchange correlations. This, however, requires intricate fine-tuning and could generally suffer from low efficiency and parasitic backscattering. Recent discovery of two-dimensional topological insulators (TI),\cite{kanePRL05,*kanePRL05z2,*bernevigPRL06,*bernevigScience06,*hasanRMP10,*qiRMP11} also called quantum spin-Hall insulators (QSHI), could provide robust means of spin-to-charge conversion, owing to its special edge states. Experimentally it is established in inverted-band HgTe quantum-well heterostructures.\cite{konigScience07,*rothScience09} The edge states of a QSHI are robust against time-reversal symmetric perturbations, and their spins and momenta are tightly correlated. A given edge of a QSHI supports a Kramers pair of counter-propagating gapless modes with opposite spins, which we call helical edge states. A CP splitter utilizing such helical edge states as charge carriers has been proposed,\cite{satoPRL10} where it was shown that the entangled spin-singlet state from CP imprints a characteristic signature in the current-current correlations. Quasi-one-dimensional semiconductor wires with strong spin-orbit coupling, such as InAs, subject to an external magnetic field can provide a way to emulate the helical states,\cite{satoPRB12} which shares many features and functionalities of helical edge states. Such a CP splitter utilizing a helical electron system was recently suggested as a mean to perform a Bell test based on nonlocal current correlations along the edges of two QSHI's.\cite{chenPRL12}

In this paper, we study a Bell test implemented by an electron-pair splitter based on the interacting helical edge states of a QSHI. Each edge state is deformed to form a beam splitter, as seen in Fig.~\ref{fig1}, replacing a spin filter in a conventional Bell-test experiment. The electron-electron interactions in the helical edge states are crucial for separating a CP into different edges of the QSHI.\cite{recherPRB02} The edge states are treated as inhomogeneous helical Luttinger liquids (LLs), whose segments in the proximity to the SC have sizable interactions, while the outside regions, which form beam splitters, are noninteracting Fermi gases. A LL wire connected to Fermi-liquid reservoirs is known to mask the effect of electron-electron interactions in ballistic transport,\cite{maslovPRB95,*safiPRB95} which simplifies the construction of a Bell inequality by the low-frequency current-current correlations. A violation of the inequality can be achieved by controlling scattering through the beam splitters via external means, such as electrostatic gating or magnetic field. At finite temperatures, the electron-electron interaction in LL leads to decoherence due to charge fractionalization,\cite{le_hurPRL05,*le_hurPRB06} suppressing signatures of the CP entanglement.

\begin{figure}[pb]
\includegraphics[width=0.8\linewidth]{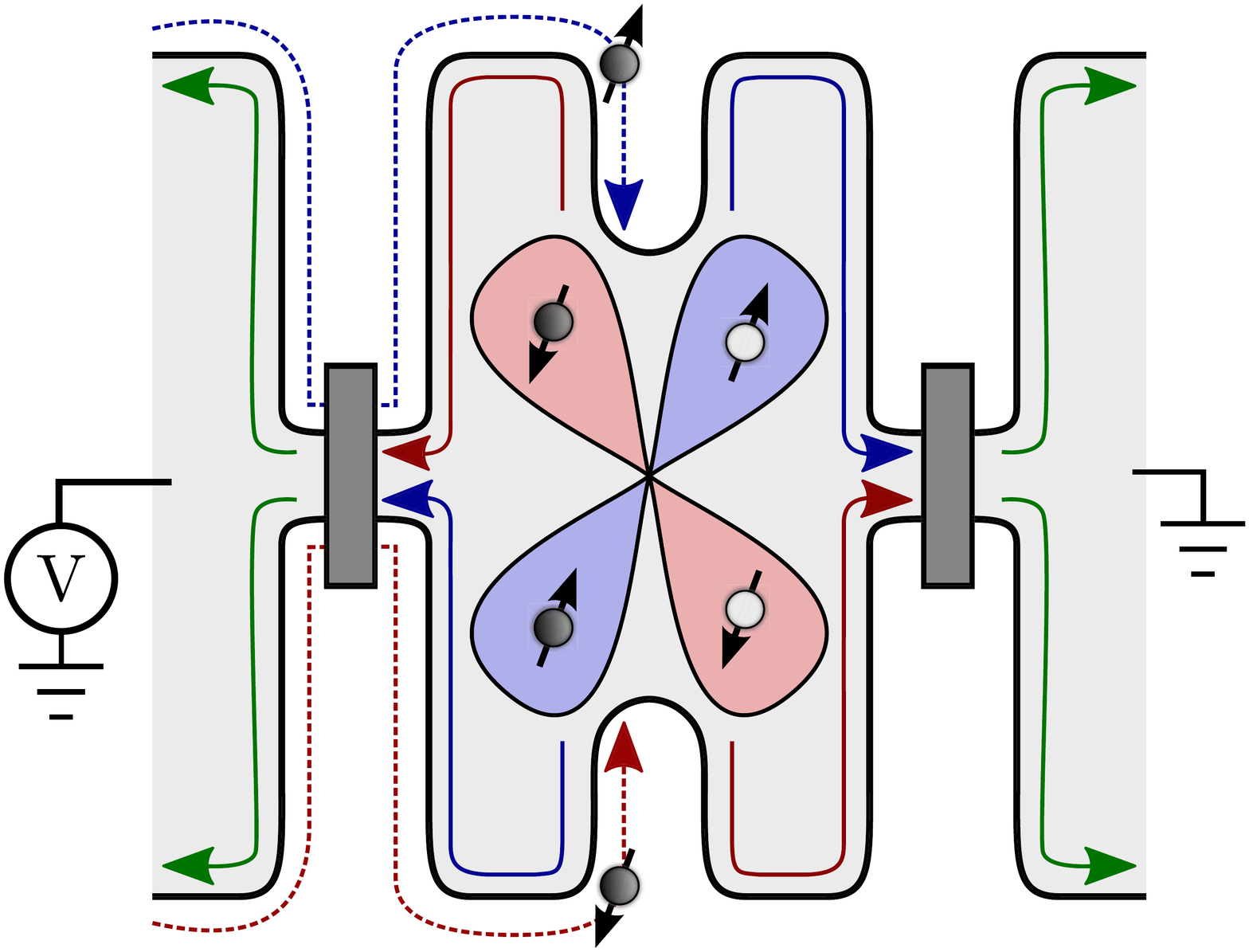}
\caption{Alternatively to the CP splitter schematically shown in Fig.~\ref{fig1}, an entangled electron-hole pair can be injected across the QSHI edges through the constriction in the middle, by biasing the left reservoir relative to the right reservoir. Blue (red) lobe indicates an electron-hole pair created by spin up (down) incoming state from the left reservoir, denoted by blue (red) dashed trajectories. These entangled electrons-hole pairs then propagate along the edges toward the two beam splitters (shaded regions).}
\label{fig2}
\end{figure}

So far, we have focused on the spin-entangled electron pairs entering different edges and going through beam splitters as in Fig.~\ref{fig1}. Alternatively, entangled electron-hole pairs can be produced via weak tunneling between the upper and lower edges analogously to Ref.~\onlinecite{beenakkerPRL03}, with the corresponding system sketched in Fig.~\ref{fig2}. The entangled electron and hole are assumed to go through the beam splitter without backscattering, and the current-current correlations at the output ports of the beam splitter can be used for constructing a Bell test.\cite{satoPRB14} As this type of system based on injecting electron-hole pairs instead of CPs can in principle be operated very similarly to the device shown in Fig.~\ref{fig1}, we will henceforth limit our discussion exclusively to the latter. The advantage of the proposals based on Figs.~\ref{fig1} or \ref{fig2} to those of Refs.~\onlinecite{samuelssonPRL03,beenakkerPRL03} is that the interedge tunneling naturally creates maximally entangled quasiparticle pairs of the form $(|{\uparrow\downarrow}\rangle\mp|{\downarrow\uparrow}\rangle)/\sqrt{2}$ (spin here corresponding to chirality of edge index).\cite{Note1} It is important to mention that in the models depicted in Figs.~\ref{fig1} and \ref{fig2}, we assume structural inversion symmetry in the central, tunneling region, in addition to the time-reversal symmetry (thus dictating effective spin conservation on tunneling\cite{satoPRL10}). In contrast, momentum conservation is assumed for the beam splitters, which requires locally lifting inversion symmetry (e.g., by a Rashba coupling) or time-reversal symmetry (e.g., by a magnetic or exchange field), in order to allow for interedge scattering.

\section{Model and Hamiltonian}

We consider a CP splitter formed by tunnel coupling a $s$-wave SC to the helical edge states of a QSHI. The total Hamiltonian of the system is $H=H_0+H_T$, where $H_0$ describes the unperturbed edge states, including electron-electron interactions, and the tunneling from the SC is given by the  Hamiltonian $H_T$.
Since the spin and momentum of the edge states are locked by their helical structure, the LL branches can be labelled by chiral index, $r=\pm$, for the right- and left-moving states, respectively (the spin index is redundant). We suppose the spin-up (down) states circulate clockwise (counterclockwise) around the QSHI sample, as sketched in Fig.~\ref{fig1}. Hence, we denote electron field operators for the right- and left-moving LL branches on the upper ($u$) and lower ($l$) edges by
\begin{equation}\label{edge state operators}
\psi_{u,\uparrow(\downarrow)}\equiv\psi_{u,\pm}\,,\quad
\psi_{l,\uparrow(\downarrow)}\equiv\psi_{l,\mp}\,,
\end{equation}
where (dropping Klein factor and the trivial phase factor $e^{irk_F x}$ associated with the Fermi wave number $k_F$)
\begin{equation}\label{fermion to boson}
\psi_{n,r}\propto\frac{e^{i(\theta_n+r\phi_n)}}{\sqrt{2\pi\delta}}\,,
\end{equation}
in terms of bosonic fields $\phi$ and $\theta$, subjected to a short-distance cutoff $\delta$.\cite{giamarchiBook} In our convention, the commutation relation for the bosonic operators is given by $\left[\theta_n(x),\phi_{n'}(y)\right]=i(\pi/2)\delta_{nn'}\text{sgn}(x-y)$. The effective LL Hamiltonian of the helical edge states in terms of the bosonic operators reads
\begin{equation}\label{H0}
H_0=\sum_{n=u,l}\int\frac{dx}{2\pi}v(x)\left[\frac{1}{g(x)}(\partial_x\phi_n)^2+g(x)(\partial_x\theta_n)^2\right]\,,
\end{equation}
where $g$ is the interaction parameter and $v$ is the renormalized velocity of the plasmonic excitations, both position dependent for an inhomogeneous LL. For both edges, we take $x=0$ as the point where electrons tunnel from the SC, and we let the interacting region be $|x|\le L$ with $g(x)=g<1$ (repulsive interaction). Exterior of this region ($|x|>L$) is noninteracting, where we set $g(x)=1$. Correspondingly, the velocity is $v(x)=v_F$ for $|x|>L$ and $v(x)=v$ for $|x|\leq L$. In addition, the left and right ends of each edge are connected through a beam splitter (see Fig. 1), which will be treated using scattering-matrix formalism.

The temperature $k_BT$ and the voltage bias $eV$ between the SC and QSHI are set below the superconducting energy gap $\Delta$ to prevent quasiparticle tunneling. We will be interested in the low-temperature regime, $k_BT\ll eV$, when the electric shot noise dominates over the thermal noise. In order to achieve CP splitting into different QSHI edges, their separation should be less than the superconducting coherence length. Furthermore, the electron-electron (here, LL) interaction is necessary to suppress the same-edge tunneling.\cite{recherPRB02} Large enough interaction strength $g$ and gap $\Delta$ allow for the different-edge tunneling to become the dominant transport process. This allows one to employ a simple model of equal-time cross-edge tunneling of spin-singlet electron pairs. As the spin-singlet wave function corresponds to $\psi_{u,\uparrow}\psi_{l,\downarrow}-\psi_{u,\downarrow}\psi_{l,\uparrow}$, one obtains the following tunneling Hamiltonian (assuming structural inversion symmetry\cite{satoPRL10}):
\begin{align}\label{tunnel}
H_T&=\Gamma e^{-i\omega_0 t}\left[\psi_{u,+}(0)\psi_{l,+}(0)-\psi_{u,-}(0)\psi_{l,-}(0)\right]+\text{H.c.}\nonumber\\
&=\sum_{\substack{r,\varepsilon=\pm}}\varepsilon r\Gamma^\varepsilon e^{-i\varepsilon\omega_0 t}\psi^\varepsilon_{u,r}(0)\psi^\varepsilon_{l,r}(0)\,.
\end{align}
Here, $\Gamma$ is a CP tunneling coefficient, $\omega_0=2eV/\hbar$ is the Josephson frequency, and $\varepsilon$ labels Hermitian conjugate: $\psi^{+}=\psi$ and $\psi^-=\psi^\dagger$.\cite{Note2}

\section{Beam Splitters}

The ends of each edge of the QSHI, where the interaction vanishes ($g=1$), are connected to a beam splitter as in Fig.~\ref{fig1}. The regions forming the beam splitters are made sufficiently long (on the scale of the Fermi wavelength), so that the momentum is effectively conserved. Hence, we assume no backscattering occurs from the beam splitters. In a given edge, the right- and left-moving incoming and outgoing states through the beam splitter are related by
\begin{equation}\label{scattering matrix}
\begin{pmatrix}
\psi_{n,+}'\\\psi_{n,-}'
\end{pmatrix}=
\begin{pmatrix}
\cos\frac{\varphi_n}{2}&-\sin\frac{\varphi_n}{2}\\
\sin\frac{\varphi_n}{2}&\cos\frac{\varphi_n}{2}\\
\end{pmatrix}
\begin{pmatrix}
\psi_{n,+}\\\psi_{n,-}
\end{pmatrix}\,,
\end{equation}
where $\psi_{n,\pm}'$ and $\psi_{n,\pm}$ refer to the right (left)-moving outgoing and incoming states, respectively, along the $n$th edge ($n=u,l$). $\varphi_n$ is the beam-splitter scattering angle, which can be controlled by local electromagnetic or elastic means.\cite{liuNature98,*oliverScience99}

The current operators at the detection points denoted by $u_\pm$ and $l_\pm$ in Fig.~\ref{fig1} can be readily expressed in terms of the outgoing filed operators $\psi_{n,r}'$. Defining the currents to be positive away from the beam splitters, the current operator $I^n_{\pm}$ at the edge $n=u,l$ for the right ($+$) and left ($-$) detection points is given by
\begin{equation}\label{current}
I^n_\pm=ev_F\psi_{n,\pm}'^\dagger\psi_{n,\pm}'
=I_{\pm,+}^n+I_{\pm,-}^n+I_{\pm,i}^n\,,
\end{equation}
where $v_F$ is the Fermi velocity in the noninteracting leads. Using Eq.~\eqref{scattering matrix}, three different terms appearing in Eq.~(\ref{current}) are given by
\begin{align}\label{current operators}
I^n_{\pm,r}&=\frac{ev_F}{2}(1\pm r\cos\varphi_n)\psi_{n,r}^\dagger\psi_{n,r}\,,\nonumber\\
I^n_{\pm,i}&=\mp\frac{ev_F}{2}\sin\varphi_n\left(\psi^\dagger_{n,+}\psi_{n,-}+\text{H.c.}\right)\,,
\end{align}
where $\psi_{n,+}$ and $\psi_{n,-}$ are evaluated at some reference points $x>L$ and $x<-L$, respectively, before reaching the beam splitters. There are two types of contributions to the currents, namely the incoherent current, $I^n_{\pm,r}$, which is insensitive to dephasing along the edges, and the interference current, $I^n_{\pm,i}$, which carries the crucial quantum-phase information.

\section{Current and Noise}

Two spin-entangled electrons initially constituting a CP are spatially separated into the top and bottom edges, with the currents produced by such entangled electrons being correlated accordingly to the edge helicity. Thus, the ensuing current-current correlations reflect the entanglement of the injected electron pair. In the following, we calculate the average current and the low-frequency noise, to the leading order in tunneling.

The expectation value of current $\bar I^n_\pm$ along the $n$th edge ($n=u,l$) is given perturbatively by
\begin{align}\label{average current initial form}
\bar I^n_\pm=&\left\langle I^{n}_\pm(t)\right\rangle=\left\langle T_c e^{-\frac{i}{\hbar}\int_c dt'H_T(t')}I^{n}_\pm(t,+)\right\rangle\nonumber\\
\approx&-\frac{1}{2\hbar^2}\sum_{ \eta_1,\eta_2=\pm}\eta_1\eta_2\int dt_1 dt_2\nonumber\\
&\times\left\langle T_c H_T(t_1,\eta_1)H_T(t_2,\eta_2)I^{n}_{\pm}(t,+)\right\rangle\,.
\end{align}
The time evolution of the operators here is given by the interaction picture by $A(t)=e^{iH_0t/\hbar}Ae^{-iH_0t/\hbar}$. $T_c$ stands for the Keldysh contour ordering and $\eta$ labels its branches, with $\eta=\pm$ for the upper (lower) branch. Using Eq.~(\ref{tunnel}) and (\ref{current operators}), the above Eq.~(\ref{average current initial form}) can be expressed in terms of the incoming fermionic operators. Following the standard bosonization scheme, we proceed by expressing the fermionic operators in terms of the bosonic operators following Eq.~(\ref{fermion to boson}), and the chiral electron density appearing in the current operators can be written as $\psi^\dagger_{n,r}\psi_{n,r}=\partial_x(\phi_n+r\theta_n)/2\pi$.\cite{giamarchiBook}

A detailed calculation for the average current is, for completeness, included in Appendix~\ref{app. current}. The final expression is given in terms of the Green's functions for bosonic fields that incorporate the appropriate boundary conditions for the inhomogeneous LL, Eqs.~(\ref{Green's functions |x|>L}) and (\ref{Green's functions |x|<L}). The current reads
\begin{equation}\label{average current}
\bar I^n_\pm
=\left\langle I^n_{\pm,+}(t)+I^n_{\pm,-}(t)\right\rangle
=\frac{e}{h}\left(\frac{|\Gamma|}{\hbar v}\right)^2P(\omega_0)\equiv\bar{I}\,,
\end{equation}
where $P(\omega)$, which is defined in terms of the Fourier transform of the function $P_{\pm\mp}(t)$, Eq.~(\ref{QP}), is proportional to the product of the tunneling densities of states of the edge LLs and independent of the beam splitter scattering angle $\varphi_n$. Note that $\left\langle I^n_{\pm,i}\right\rangle=0$ for our forward-scattering beam splitter.

As an electron tunnels into the interacting region ($|x|\leq L$), resulting plasmonic charge-density waves go trough multiple reflections between the interfaces of interacting and noninteracting regions (at $x=\pm L$), for which the interacting region acts as a Fabry-P\'{e}rot resonator.\cite{safiPRB95} Such reflections are seen as multiple oscillations in the bosonic Green's functions, as in Eqs.~(\ref{Green's functions |x|>L}) and (\ref{Green's functions |x|<L}), and the fermionic Green's functions oscillate in turn. The propagation time $t_L=L/v$ of the plasmonic excitations across the interacting region sets the time scale of the Fabry-P\'{e}rot oscillation. The function $P_{\pm\mp}(t)$ is a product of the part related to the Green's function in the absence of noninteracting leads (i.e., $L\to\infty$) and the factor containing the effect of the Fabry-P\'{e}rot resonator. The applied bias $V$ sets the time scale $t_V=\hbar/eV=2\omega_0^{-1}$. When $t_V\ll t_L$ (large bias), the phase $e^{i\omega_0 t}$ in the Fourier transform of $P_{\pm\mp}(t)$ oscillates more rapidly than the time scale of the Fabry-P\'{e}rot oscillation. In this limit, the effect of the resonator is washed out, and we can evaluate $P(\omega_0)$ in the absence of the noninteracting leads,\cite{lebedevPRB05} finding $P(\omega_0)\propto\omega_0^{2\gamma+1}$. Here, $\gamma=(g+g^{-1}-2)/2$ is the single-particle tunneling density of states exponent in a bulk LL.

The symmetrized current-current correlators between the upper and lower edges are given by
\begin{align}\label{noise}
S_{\alpha\beta}(t,t')&=\left\langle\{\delta I^{u}_\alpha(t),\delta I^{l}_\beta(t')\}\right\rangle\nonumber\\
&=\sum_{\eta=\pm}\left\langle T_ce^{-\frac{i}{\hbar}\int_c dt''H_T(t'')}I^{u}_\alpha(t,\eta)I^{l}_\beta(t',-\eta)\right\rangle\,,
\end{align}
where $\delta I^n_\alpha(t)=I^n_\alpha(t)-\bar I$ is the current fluctuation. The above correlation is evaluated up to second order in $\Gamma$. The current correlations come in various combinations the incoherent currents $I^n_{\alpha,+}$ and $I^n_{\alpha,-}$, and the interference current $I^n_{\alpha,i}$. Let us decompose the noise, $S_{\alpha\beta}(t,t')=\sum_{\mu,\nu=\pm,i}S_{\alpha\beta}^{\mu\nu}(t,t')$, into terms corresponding to different current combinations of the upper-edge current $I^u_{\alpha,\mu}$ and the lower-edge current $I^l_{\beta,\nu}$:
\begin{align}
&S_{\alpha\beta}^{\mu\nu}(t,t')\nonumber\\
&=\sum_{\eta=\pm}\left\langle T_ce^{-\frac{i}{\hbar}\int_c dt''H_T(t'')}I^{u}_{\alpha,\mu}(t,\eta)I^{l}_{\beta,\nu}(t',-\eta)\right\rangle\,.
\end{align}
The cross terms between the interference part $I_{\alpha,i}$ and the incoherent part $I_{\alpha,\pm}$ give no contribution. The terms involving only the incoherent current $I^n_{\alpha,\pm}$ result in
\begin{equation}\label{Spm}
\tilde S_{\alpha\beta}^{(0)}\equiv\sum_{\mu,\nu=\pm}\tilde S_{\alpha\beta}^{\mu\nu}=
e\bar I\left(1+\alpha\beta\cos\varphi_u\cos\varphi_l\right)\,.
\end{equation}
Here, $\tilde S^{\mu\nu}_{\alpha\beta}\equiv S^{\mu\nu}_{\alpha\beta}(\omega=0)$ is the zero-frequency Fourier transform of $S^{\mu\nu}_{\alpha\beta}(t-t')$. Lastly, we find the correlation involving only the interference terms $I^n_{\alpha,i}$ as
\begin{equation}\label{S00}
\tilde S^{(i)}_{\alpha\beta}=\alpha\beta C(\omega_0)e\bar I\sin\varphi_u\sin\varphi_l\,.
\end{equation}
$C(\omega)$ is the Fourier transform of $C(t)$ given in Eq.~(\ref{C}). It characterizes dephasing and ranges $0\le C(\omega_0)\le 1$. When $g=1$ (i.e., the edges are everywhere noninteracting), $C(\omega_0)=1$, which means the nonlocal spin entanglement of the electron pair persists until the currents are measured. In this ideal case, the total noise is given by (for $r=\pm$)
\begin{equation}\label{S ideal}
\tilde S_{r,\pm r}=e\bar I\left[1\pm\cos(\varphi_u-\varphi_l)\right]\,.
\end{equation}
This form of noise reminds us of the spin correlations in the EPR thought experiment, where a spin-singlet state decays into two counter-propagating particles, whose resulting beams pass through two distant polarizers before being detected. The coincidence signal correlations in the distant detectors depend sinusoidally on the relative angle of the polarizers. In Eq.~(\ref{S ideal}), our current correlations similarly depend on the relative scattering angle of the beam splitters.

LL is known to exhibit a charge fractionalization,\cite{phamPRB00,*lehurAP08,*steinbergNPhys08} where a chiral single-particle state, say a right-moving electron, breaks down into a charge $e(1+g)/2$ moving to the right and $e(1-g)/2$ moving to the left. At finite temperature, these counter-propagating states cease to overlap after a time $\tau=\pi k_BT\gamma/\hbar$, as is reflected in the exponential decay (dephasing) of a single-particle propagator for the right-moving branch.\cite{le_hurPRL05,*le_hurPRB06} The interference effect is likewise exponentially suppressed. For instance, exponential suppression in the Aharonov-Bohm oscillation of the tunneling current between two LL wires has been studied in Ref.~\onlinecite{le_hurPRL05}.

When the electron temperature is above the finite-size crossover temperature, $T\gg T_0\equiv\hbar v/2\pi k_BL$, the interference in an LL system of size $2L$ decays exponentially. If $T\ll T_0$, the suppression occurs in a power-law form in a complicated fashion depending on the hierarchy of the relevant energy scales: ambient temperature, $k_BT$, bias, $eV$, and the crossover temperature, $k_BT_0$. In our case, this dephasing affects $C$ appearing in Eq.~(\ref{S00}), which is expected to show similar reduction at finite temperatures. Using the Green's functions in Eqs.~(\ref{Green's functions |x|>L}) and (\ref{Green's functions |x|<L}), we can extract the exponentially decaying part, which is given by $C(\omega_0)\propto e^{-2\gamma T/T_0}$. Such exponential suppression does not affect $\tilde S_{\alpha\beta}^{(0)}$ pertaining to the incoherent current, as a consequence of the conservation of charge. In low-temperature regime, $T\ll T_0$, $C(\omega_0)$ is instead expected to show a power-law behavior, with details depending on the relative strength of the bias with respect to the crossover energy scale (i.e., $k_BT\ll eV\ll k_BT_0$ or $k_BT\ll k_BT_0\ll eV$).\cite{gellerPRB97,le_hurPRB06}

\section{Bell inequality}

In optical experiments, a violation of a Bell inequality is tested by coincidence counting of the simultaneous arrival of a pair of entangled photons at remote locations. On the other hand, it is more natural to measure current correlations in solid-state devices, which could be used to construct a Bell inequality in beam-splitter based systems.\cite{kawabataJPSJ01,*chtchelkatchevPRB02} The time window for a current measurement should be short enough so that no more than a single Cooper pair is detected at a time and the $1/f$ noise can be neglected, but it should also be sufficiently long on the scale of the inverse voltage and the transport time along the edges such that the zero-frequency approximation for the shot noise is adequate.\cite{chtchelkatchevPRB02} Under these conditions, the current-current correlations can be combined to give the Clauser-Horne-Shimony-Holt Bell inequality.\cite{clauserPRL69} As shown in the previous section, the total zero-frequency noise is (evaluating $C$ at $\omega_0$ throughout)
\begin{equation}
\tilde{S}_{\alpha\beta}
=e\bar{I}\left[1+\alpha\beta\left(\cos\varphi_u\cos\varphi_l+C\sin\varphi_u\sin\varphi_l\right)\right]\,.
\end{equation}
The Bell inequality then is given by
\begin{align}\label{Bell inequality}
B&\equiv|E(\varphi_u,\varphi_l)-E(\varphi_u,\varphi_l')+E(\varphi_u',\varphi_l)+E(\varphi_u',\varphi_l')|\nonumber\\
&\leq 2\,,
\end{align}
where the correlation functions in the inequality are directly related to the noise spectra by
\begin{align}\label{E}
E(\varphi_u,\varphi_l)&=\frac{\tilde{S}_{++}-\tilde{S}_{+-}-\tilde{S}_{-+}+\tilde{S}_{--}}{\tilde{S}_{++}+\tilde{S}_{+-}+\tilde{S}_{-+}+\tilde{S}_{--}}\nonumber\\
&=\cos\varphi_u\cos\varphi_l+C\sin\varphi_u\sin\varphi_l\,.
\end{align}
The noninteracting ($g=1$) zero-temperature case gives maximally-entangled result with $C=1$ and $E(\varphi_u,\varphi_l)=\cos(\varphi_u-\varphi_l)$. A choice of the angles maximizing $B$ is $\varphi_u=\pi/4$, $\varphi_l=\pi/2$, $\varphi_u'=3\pi/4$, and $\varphi_l'=\pi$, leading to $B=2\sqrt 2$.

Even in the presence of dephasing, i.e., $C<1$, by adjusting the four angles, $\varphi_u$, $\varphi_u'$, $\varphi_l$, and $\varphi_l'$, the maximum value of the Bell parameter\cite{samuelssonPRL03}
\begin{equation}
B=2\sqrt{1+C^2}
\end{equation}
still exceeds $2$. This means that the Bell inequality can in principle be violated for arbitrary nonzero $C$. The optimal violation angles are given by\cite{samuelssonPRL03}
\begin{align}
&\tan\varphi_u=-C\cot\varphi_s\,,\quad
\tan\varphi_u'=C\tan\varphi_s\,,\nonumber\\
&\tan\frac{\varphi_l-\varphi_l'}{2}=\text{sgn}(\cos\varphi_u)\sqrt{\frac{\tan^2\varphi_s+C^2}{C^2\tan^2\varphi_s+1}}\,,
\end{align}
where $\varphi_s\equiv(\varphi_l+\varphi_l')/2$ is arbitrary. Although it is possible to observe a violation of the Bell inequality under a finite dephasing, the range of angles that can achieve a violation shrinks as $C\to0$.

\section{Discussion and Conclusion}

We discussed the construction of a Bell inequality via the current-current correlations between different edges of a QSHI equipped with beam splitters. The entanglement is produced by coherently injecting electron Cooper-pairs from a superconductor or electron-hole pairs from a normal Fermi-liquid reservoir biased by a constant voltage $V$ with respect to the QSHI. Adjusting the transmission matrix through the beam splitters by local electric or magnetic fields, a violation of the Bell inequality can be achieved even in the presence of a moderate dephasing, parametrized by $C$ (with $C=1$ corresponding to maximal entanglement with no dephasing and $C=0$ to complete dephasing and classical correlations).

The edge states of a QSHI are modeled as helical LLs. Electron-electron interactions are essential ingredients in order to achieve tunneling of two electrons forming a CP into different edges. On the other hand, the charge fractionalization furnished by LL causes dephasing at finite temperature when $T>T_0=\hbar v/2\pi k_B2L$. In this high-temperature regime, the dephasing parameter suffers exponential decay as $C\propto e^{-2\gamma T/T_0}$. In the low-temperature limit, $T<T_0$, $C$ does not decay exponentially, but is expected to follow power-law scaling characteristic of LLs. Even with the reduction of the dephasing parameter below unity, the entanglement of quasiparticle (electron-electron or electron-hole) pairs is visible through the violation of the Bell inequality, albeit it becomes progressively more difficult to tune the beam splitters to achieve the violation as $C$ vanishes.

The QSHI edge states thus provide a promising medium for production and manipulation of quantum information in mesoscopic systems, even in the absence of any correlations (as in Fig.~\ref{fig2}). In our minimal model, we have only considered dephasing due to internal electronic interactions along the edges. Collective or quasiparticle modes present in the solid-state environment can generally be expected to provide additional detrimental dephasing sources that need to be studied and mitigated.

\acknowledgments

This work was supported by the NSF under Grant No. DMR-0840965. We gratefully acknowledge fruitful discussions with Daniel Loss and Mircea Trif.

\appendix

\section{Green's functions}

Evaluation of the current and noise in Eqs.~(\ref{average current initial form}) and (\ref{noise}) is based on several bosonic Green's functions. Since the system of interest here is an inhomogeneous LL where the interaction parameter $g$ depends on the position, we need to impose appropriate boundary conditions to obtain the Green's functions.

First, we identify the Lagrangian for the bosonic fields $\phi$ and $\theta$ from Eq.~(\ref{H0}) as
\begin{equation}\label{full lagrangian}
\mathcal{L}
=\frac{1}{\pi}\partial_x\theta\partial_t\phi-\frac{v}{2\pi}\left[\frac{1}{g}(\partial_x\phi)^2+g(\partial_x\theta)^2\right]\,.
\end{equation}
The effective Lagrangian for the $\phi$ or $\theta$ field can be found by integrating out the $\theta$ or $\phi$ field, respectively:
\begin{align}\label{lagrangian}
\mathcal{L}_{\phi}&=\frac{1}{2\pi g}\left[\frac{1}{v}(\partial_t\phi)^2-v(\partial_x\phi)^2\right]\,,\nonumber\\
\mathcal{L}_{\theta}&=\frac{g}{2\pi}\left[\frac{1}{v}(\partial_t\theta)^2-v(\partial_x\theta)^2\right]\,.
\end{align}
The spatial dependence of the velocity and the interaction parameter are $v(x)=v_F$ and $g(x)=g_l=1$ for $|x|>L$, and $v(x)=v$ and $g(x)=g$ for $|x|\leq L$. For electrons injected at $x=0$, the retarded Green's functions are found to satisfy the following differential equations:
\begin{align}
&\dfrac{1}{\pi}\left[\frac{\omega^2}{g(x)v(x)}+\partial_x\left(\frac{v(x)}{g(x)}\partial_x\right)\right]
G_R^{\phi\phi}(x,\omega)=\delta(x)\,,\nonumber\\
&\dfrac{1}{\pi}\left[\frac{g(x)\omega^2}{v(x)}+\partial_x\left(v(x)g(x)\partial_x\right)\right]G_R^{\theta\theta}(x,\omega)=\delta(x)\,,
\end{align}
with the appropriate boundary conditions: $(1)$ the solutions in the leads are moving away from $x=0$, $(2)$ $G_R(x,\omega)$ is continuous at $x=\pm L, 0$, $(3)$ the following expressions at $x=\pm L$ are continuous:\cite{Note3}
\begin{align*}
\frac{v(x)}{g(x)}\partial_xG_R^{\phi\phi}(x,\omega)\bigg\vert_{x=\pm L+0^-}^{x=\pm L+0^+}&=0\,,\\
v(x)g(x)\partial_xG_R^{\theta\theta}(x,\omega)\bigg\vert_{x=\pm L+0^-}^{x=\pm L+0^+}&=0\,,
\end{align*}
and (4) the derivative at the location of the delta function $x=0$ is discontinuous as
\begin{align*}
\frac{v(x)}{g(x)}\partial_xG_R^{\phi\phi}(x,\omega)\bigg\vert_{x=0^-}^{x=0^+}&=\pi\,,\\
v(x)g(x)\partial_xG_R^{\theta\theta}(x,\omega)\bigg\vert_{x=0^-}^{x=0^+}&=\pi\,.
\end{align*}

\begin{widetext}
We look for the solutions of the form
\begin{equation}
G_R^{\phi\phi,\theta\theta}(x,\omega)=\left\{
\begin{array}{ll}
Ae^{-i\omega x/v_F} & {\rm for~}x<-L\\
Be^{i\omega x/v}+Ce^{-i\omega x/v} & {\rm for~}-L\le x\le 0\\
De^{i\omega x/v}+Ee^{-i\omega x/v} & {\rm for~}0\le x \le L\\
Fe^{i\omega x/v_F} & {\rm for~}L<x\\
\end{array}
\right.\,,
\end{equation}
which, after imposing the above boundary conditions, we find
\begin{align}\label{G_retarded}
G_R^{\phi\phi}(x,\omega)
&=A_\phi g
\left\{
\begin{array}{ll}
a_te^{i\omega(|x|-L)/v_F} & {\rm for~}L<|x|\\
e^{i\omega(|x|-L)/v}+a_re^{-i\omega(|x|-L)/v} & {\rm for~}|x| \le L\\
\end{array}
\right.\,,\,\,\,A_\phi=-i\frac{\pi}{2\omega}
\frac{e^{i\omega L/v}}{1-a_re^{i2\omega L/v}}\,,\nonumber\\
G_R^{\theta\theta}(x,\omega)
&=A_\theta\left\{
\begin{array}{ll}
g_l^{-1}a_te^{i\omega(|x|-L)/v_F} & {\rm for~}L<|x|\\
g^{-1}e^{i\omega(|x|-L)/v}-a_re^{-i\omega(|x|-L)/v} & {\rm for~}|x| \le L\\
\end{array}
\right.\,,\,\,\,A_\theta=-i\frac{\pi}{2\omega}
\frac{e^{i\omega L/v}}{1+a_re^{i2\omega L/v}}\,.
\end{align}
\end{widetext}
Here, $a_t=2g_l/(g_l+g)$ and $a_r=(g_l-g)/(g_l+g)$ are the transmission and reflection coefficients for the bosonic fields between regions with different interaction parameter strengths. Given the retarded Green's functions, the greater and lesser Green's functions can be found by the standard relationships
\begin{equation}
G_>=i2[1+n_B(\omega)]\text{Im}G_R\,,\,\,\,G_<=i2n_B(\omega)\text{Im}G_R\,,
\end{equation}
where $n_B(\omega)=1/(e^{\beta\hbar\omega}-1)$ is the bosonic distribution function.

In the calculation of the current and noise, we encounter the Keldysh contour ordered Green's functions. The following conventions are used: $G^{AB}_{\eta\eta'}(x,t)=\left\langle T_cA(x,t,\eta)B(0,0,\eta')\right\rangle$, where
\begin{align}
G^{AB}_{-+}(x,t)&=G^{AB}_>(x,t)=-i\left\langle A(x,t)B(0,0)\right\rangle\nonumber\,,\\
G^{AB}_{+-}(x,t)&=G^{AB}_<(x,t)=-i\left\langle B(0,0)A(x,t)\right\rangle\nonumber\,,\\
G_{++}^{AB}(x,t)&=\Theta(t)G_>(x,t)+\Theta(-t)G_<(x,t)\nonumber\,,\\
G_{--}^{AB}(x,t)&=-\Theta(-t)G_>(x,t)-\Theta(t)G_<(x,t)\,,
\end{align}
for arbitrary bosonic operators $A$ and $B$.

The finite-temperature Green's functions at $|x|>L$ (noninteracting region) are found to be
\begin{widetext}
\begin{align}\label{Green's functions |x|>L}
iG^{\phi\phi}_{\eta\eta'}(x,t)&=\left\langle T_c\phi(x,t,\eta)\phi(0,0,\eta')\right\rangle
\rightarrow-\frac{g}{4}a_t\sum_{n=0}^\infty a_r^n\sum_{s=\pm}
G_{n,s}(x,t)\,,\nonumber\\
iG^{\theta\theta}_{\eta\eta'}(x,t)&=\left\langle T_c\phi(x,t,\eta)\phi(0,0,\eta')\right\rangle
\rightarrow-\frac{1}{4}a_t\sum_{n=0}^\infty (-a_r)^n\sum_{s=\pm}
G_{n,s}(x,t)\,,\nonumber\\
iG^{\theta\phi}_{\eta\eta'}(x,t)&=\left\langle\theta(x,t,\eta)\phi(0,0,\eta')\right\rangle
\rightarrow\text{sgn}(x)\frac{g}{4}a_t\sum_{n=0}^\infty a_r^n\sum_{s=\pm}s
G_{n,s}(x,t)\,,\nonumber\\
iG^{\phi\theta}_{\eta\eta'}(x,t)&=\left\langle\phi(x,t,\eta)\theta(0,0,\eta')\right\rangle
\rightarrow\text{sgn}(x)\frac{1}{4}a_t\sum_{n=0}^\infty (-a_r)^n\sum_{s=\pm}s
G_{n,s}(x,t)\,,\nonumber\\
\end{align}
where $D_{\eta\eta'}(t)=\Theta(\eta\eta')\text{sgn}(\eta't)+\Theta(-\eta\eta')\text{sgn}(\eta')$ and
\begin{equation}
G_{n,s}(x,t)=\ln\sin\left\{\frac{\pi}{\hbar\beta}\left[\frac{\delta}{v}+iD_{\eta\eta'}(t)\left(t-s\frac{L(2n+1)}{v}-s\frac{|x|-L}{v_F}\right)\right]\right\}\,.
\end{equation}
The arrow in the above equations indicate that the divergent terms on the right hand side are left out, since they can be regularized out. From the Lagrangian in Eq.~(\ref{full lagrangian}), the first two Green's functions are related to the last two by
\begin{align}\label{theta_phi}
G^{\theta\phi}_{>,<}(x,\omega)&=i\left\langle T_c\theta(x,t,\mp)\phi(0,0,\pm)\right\rangle=-i\frac{v}{g\omega}\partial_x G^{\phi\phi}_{>,<}(x,\omega)\nonumber\,,\\
G^{\phi\theta}_{>,<}(x,\omega)&=i\left\langle T_c\phi(x,t,\mp)\theta(0,0,\pm)\right\rangle=-i\frac{vg}{\omega}\partial_x G^{\theta\theta}_{>,<}(x,\omega)\,.
\end{align}

We further need the Green's functions for $x=0$ case, which are given by
\begin{align}\label{Green's functions |x|<L}
iG^{\phi\phi}_{\eta\eta'}(x=0,t)
&\rightarrow-\frac{g}{2}
\ln\sin\left[\frac{\pi}{\hbar\beta}\left(\frac{\delta}{v}+iD_{\eta\eta'}(t)t\right)\right]
-\frac{g}{2}\sum_{n=1}^\infty \sum_{s=\pm}a_r^n
\ln\sin\left[\frac{\pi}{\hbar\beta}\left(\frac{\delta}{v}+iD_{\eta\eta'}(t)\left(t+s n\frac{2L}{v}\right)\right)\right]\,,\nonumber\\
iG^{\theta\theta}_{\eta\eta'}(x=0,t)
&\rightarrow-\frac{1}{2g}\ln\sin\left[\frac{\pi}{\hbar\beta}\left(\frac{\delta}{v}+iD_{\eta\eta'}(t)t\right)\right]
-\frac{1}{2g}\sum_{n=1}^\infty \sum_{s=\pm}(-a_r)^n
\ln\sin\left[\frac{\pi}{\hbar\beta}\left(\frac{\delta}{v}+iD_{\eta\eta'}(t)\left(t+s n\frac{2L}{v}\right)\right)\right]\,.
\end{align}
\end{widetext}
By Eqs.~(\ref{G_retarded}), and (\ref{theta_phi}), we can show $G^{\theta\phi}_{\eta\eta'}(0,t)=G^{\phi\theta}_{\eta\eta'}(0,t)=0$.

A bosonic mode created at $x=0$ propagates in the interacting region, $|x|<L$, before it hits the boundary between the interacting and noninteracting regions at $x=\pm L$. Some part of the wave is transmitted into the noninteracting region, whereas the rest is reflected back into the interacting region. This process of transmission and reflection is repeated, establishing a Fabry-P\'erot resonator structure. The above Green's functions are in the form of the sum of these transmitted and reflected parts.

\section{Current}\label{app. current}
\begin{widetext}
The average current in Eq.~(\ref{average current}) up to second order in the tunneling coefficient in terms of the fermionic fields is given by
\begin{align}
\langle I^{n}_{\pm}(t)\rangle=&\langle T_c e^{-\frac{i}{\hbar}\int_c dt''H_T(t'')}I^{n}(x,t,+)\rangle\approx\frac{|\Gamma|^2}{2\hbar^2}\sum_{\eta_1,\eta_2,\varepsilon,\sigma=\pm}
\eta_1\eta_2\int_{-\infty}^\infty dt_1\int_{-\infty}^\infty dt_2e^{-i\varepsilon\omega_0(t_1-t_2)}\nonumber\\
&\times\langle T_c\psi^{\varepsilon}_{u,\sigma}(0,t_1,\eta_1)\psi^{\varepsilon}_{l,\sigma}(0,t_1,\eta_1)
\psi^{-\varepsilon}_{u,\sigma}(0,t_2,\eta_2)\psi^{-\varepsilon}_{l,\sigma}(0,t_2,\eta_2)I^{n}_{\pm}(t,+)\rangle\,.
\end{align}
Here, the fermionic operator is  $\psi_{n,r}=e^{i(\theta_n+r\phi_n)}/\sqrt{2\pi\delta}$, where $\theta_n$ and $\phi_n$ are boson fields given in Eq.~(\ref{H0}) with the commutation relation $[\theta_n(x),\phi_{n'}(y)]=i(\pi/2)\delta_{nn'}\text{sgn}(x-y)$. $r=+ (-)$ labels the right-(left-)moving state. The incoherent parts of the current operator $I_{\pm,r}^n$ ($r=\pm$) in Eq.~(\ref{current operators}) involve fermionic operators in the combination $\psi^\dagger_{n,r}\psi_{n,r}$, which can be expressed in terms of bosonic operators as
\begin{equation}
\psi^\dagger_{n,r}(x,t)\psi_{n,r}(x,t)=\frac{1}{2\pi}\partial_x[r\theta_n(x,t)+\phi_n(x,t)]
=\frac{1}{2\pi}\partial_x(-i\partial_\lambda) e^{i\lambda[r\theta_n(x,t)+\phi_n(x,t)]}\bigg\vert_{\lambda=0}\,.
\end{equation}

The expectation value of the interference current $I_{\pm,i}^n$  vanishes, since there are always operators that cannot be contracted. By summing the contributions from $I_{\pm,+}^n$ and $I_{\pm,-}^n$, the following result is obtained:
\begin{align}
\bar I=\left\langle I^{n}_{\pm}(t)\right\rangle
=&-\frac{e}{h}\frac{v_F|\Gamma|^2}{8\pi\hbar^2v^2}
\sum_{r,\varepsilon,\sigma,\eta_1,\eta_2=\pm}(1\pm r\cos\varphi_n)\eta_1\eta_2\varepsilon\nonumber\\
&\times\left[r\tilde Q^\theta_{+\eta_1,\sigma}(r)-r\tilde Q^\theta_{+\eta_2,\sigma}(r)+\sigma\tilde Q^\phi_{+\eta_1,\sigma}(r)-\sigma\tilde Q^\phi_{+\eta_2,\sigma}(r)\right]P_{\eta_1\eta_2}(-\varepsilon\omega_0)\nonumber\\
=&\frac{e}{h}\left(\frac{|\Gamma|}{\hbar v}\right)^2[P_{-+}(\omega_0)-P_{+-}(-\omega_0)]
=\text{sgn}(\omega_0)\frac{e}{h}\left(\frac{|\Gamma|}{\hbar v}\right)^2P(\omega_0)\,.
\end{align}
Here, $\tilde Q_{\eta\eta';\sigma}(r)\equiv Q_{\eta\eta';\sigma}(x,\omega=0)$, which depend only on $r\equiv{\rm sgn}(x)$. The corresponding real-time expressions for $Q$ and $P$ are given by
\begin{align}\label{QP}
Q^\theta_{\eta\eta';\sigma}(x,t)&=\partial_xG^{\theta\theta}_{\eta\eta'}(x,t)+\sigma \partial_xG^{\theta\phi}_{\eta\eta'}(x,t)\,,\,\,\,Q^\phi_{\eta\eta';\sigma}(x,t)=\partial_xG^{\phi\phi}_{\eta\eta'}(x,t)+\sigma \partial_xG^{\phi\theta}_{\eta\eta'}(x,t)\,,\nonumber\\
&P_{\eta_1\eta_2}(t)=\frac{h v^2}{(2\pi\delta)^2}e^{
i2\left[G^{\theta\theta}_{\eta_1\eta_2}(0,t)+G^{\phi\phi}_{\eta_1\eta_2}(0,t)-G^{\theta\theta}(0,0)-G^{\phi\phi}(0,0)\right]}\,.
\end{align}
We can show that $P_{\mp\pm}(\omega)=\Theta(\pm\omega)P(\omega)$, hence $P_{-+}(\omega_0)-P_{+-}(\omega_0)=\text{sgn}(\omega_0)P(\omega_0)$. Furthermore, the relations
\begin{equation}
\tilde Q_{++,\sigma}^{\theta/\phi}(x)-\tilde Q_{+-,\sigma}^{\theta/\phi}(x)=\tilde Q_{-+,\sigma}^{\theta/\phi}(x)-\tilde Q_{--,\sigma}^{\theta/\phi}(x)=\frac{\pi}{2v_F}[\text{sgn}(x)+\sigma]
\end{equation}
turn out to be independent of temperature.
\end{widetext}
\section{Zero-frequency noise}\label{app. noise}
With the current in Eq.~(\ref{current operators}) and tunneling Hamiltonian in Eq.~(\ref{tunnel}), the current-current correlations between the upper and lower edges Eq.~(\ref{noise}), up to second order in the tunneling coefficient, are given by
\begin{widetext}
\begin{align}
S_{\alpha\beta}(t,t')=\sum_{\mu,\nu=\pm,i}S^{\mu\nu}_{\alpha\beta}(t,t')
=&-\frac{1}{2\hbar^2}\sum_{\mu,\nu=\pm,i}~\sum_{\eta,\eta_1,\eta_2,\sigma_1,\sigma_2,\varepsilon_1,\varepsilon_2=\pm}
\Gamma^{\varepsilon_1}\Gamma^{\varepsilon_2}\varepsilon_1\varepsilon_2 \sigma_1\sigma_2
\eta_1\eta_2\int_{-\infty}^\infty dt_1\int_{-\infty}^\infty dt_2e^{-i\omega_0(\varepsilon_1t_1+\varepsilon_2t_2)}\nonumber\\
&\times\langle T_c I^{u}_{\alpha,\mu}(t,\eta)I^{l}_{\beta,\nu}(t',-\eta)
\psi^{\varepsilon_1}_{u,\sigma_1}(0,t_1,\eta_1)\psi^{\varepsilon_1}_{l,\sigma_1}(0,t_1,\eta_1) \psi^{\varepsilon_2}_{u,\sigma_2}(0,t_2,\eta_2)\psi^{\varepsilon_2}_{l,\sigma_2}(0,t_2,\eta_2)\rangle\,,
\end{align}
where $S_{\alpha\beta}^{\mu\nu}$ is the correlation between the currents $I^u_{\alpha,\mu}$ and $I^l_{\beta,\nu}$.

First, we calculate the contributions from $I_{\alpha,+}^n$ and $I_{\alpha,-}^n$:
\begin{align}
\sum_{\mu,\nu=\pm}&S^{\mu\nu}_{\alpha\beta}(t,t')\\
=&\frac{|\Gamma|^2}{2\hbar^2}\sum_{\mu,\nu=\pm}~\sum_{\eta,\eta_1,\eta_2,\sigma,\varepsilon=\pm}
\eta_1\eta_2\int_{-\infty}^\infty dt_1\int_{-\infty}^\infty dt_2e^{-i\varepsilon\omega_0(t_1-t_2)}\nonumber\\
&\times\langle T_c I^{u}_{\alpha,\mu}(t,\eta)I^{l}_{\beta,\nu}(t',-\eta)
\psi^{\varepsilon}_{u,\sigma}(0,t_1,\eta_1)\psi^{\varepsilon}_{l,\sigma}(0,t_1,\eta_1) \psi^{-\varepsilon}_{u,\sigma}(0,t_2,\eta_2)\psi^{-\varepsilon}_{l,\sigma}(0,t_2,\eta_2)\rangle\nonumber\\
=&\frac{|\Gamma|^2}{2\hbar^2}\left(\frac{ev_F}{2\pi}\right)^2
\sum_{\mu,\nu=\pm}
\frac{1}{4}(1+\alpha\mu\cos\varphi_u)(1+\beta\nu\cos\varphi_l)
\sum_{\eta,\eta_1,\eta_2,\sigma,\varepsilon=\pm}
\eta_1\eta_2\int_{-\infty}^\infty dt_1\int_{-\infty}^\infty dt_2e^{-i\varepsilon\omega_0(t_1-t_2)}\nonumber\\
&\times\left(\frac{1}{2\pi\delta}\right)^2\partial_x\partial_{x'}\langle T_c [\mu\theta_u(x,t,\eta)+\phi_u(x,t,\eta)][\nu\theta_l(x',t',-\eta)+\phi_l(x',t',-\eta)]\nonumber\\
&e^{i\varepsilon[\theta_u(0,t_1,\eta_1)+\sigma\phi_u(0,t_1,\eta_1)]}
e^{i\varepsilon[\theta_l(0,t_1,\eta_1)+\sigma\phi_l(0,t_1,\eta_1)]}
e^{-i\varepsilon[\theta_u(0,t_2,\eta_2)+\sigma\phi_u(0,t_2,\eta_2)]}
e^{i\varepsilon[\theta_l(0,t_2,\eta_2)+\sigma\phi_l(0,t_2,\eta_2)]}\big\vert{\substack{x\rightarrow \mu\\x'\rightarrow \nu}}\nonumber\\
=&\frac{|\Gamma|^2}{2\hbar^2}\left(\frac{ev_F}{2\pi}\right)^2\frac{1}{4hv^2}
\sum_{\mu,\nu=\pm} (1+\alpha\mu\cos\varphi_u)(1+\beta\nu\cos\varphi_l)
\sum_{\eta,\eta_1,\eta_2,\sigma,\varepsilon=\pm}
\eta_1\eta_2\int_{-\infty}^\infty dt_1\int_{-\infty}^\infty dt_2e^{-i\varepsilon\omega_0(t_1-t_2)}\nonumber\\
&\times\left[\mu Q^\theta_{\eta\eta_1,\sigma}(\mu,t-t_1)-\mu Q^\theta_{\eta\eta_2,\sigma}(\mu,t-t_2)
+\sigma Q^\phi_{\eta\eta_1,\sigma}(\mu,t-t_1)-\sigma Q^\phi_{\eta\eta_2,\sigma}(\mu,t-t_2)\right]\nonumber\\
&\times\left[\nu Q^\theta_{-\eta\eta_1,\sigma}(\nu,t'-t_1)-\nu Q^\theta_{-\eta\eta_2,\sigma}(\nu,t'-t_2)
+\sigma Q^\phi_{-\eta\eta_1,\sigma}(\nu,t'-t_1)-\sigma Q^\phi_{-\eta\eta_2,\sigma}(\nu,t'-t_2)\right]P_{\eta_1\eta_2}(t_1-t_2)\nonumber
\end{align}
Here, $Q^{\theta/\phi}_{\eta\eta'}(x,t)$ and $P_{\eta\eta'}(t)$ are defined in Eq.~(\ref{QP}). The zero-frequency component of the Fourier transform of the above expression is given by
\begin{align}\label{S_RL final}
\sum_{\mu,\nu=\pm}\tilde S^{\mu\nu}_{\alpha\beta}
&=\frac{e^2}{h}\frac{|\Gamma|^2}{2\hbar^2v^2}\frac{v_F^2}{4\pi^2}
\frac{1}{4}\sum_{\mu,\nu=\pm} (1+\alpha\mu\cos\varphi_u)(1+\beta\nu\cos\varphi_l)\sum_{\eta,\eta_1,\eta_2,\sigma,\varepsilon=\pm}\eta_1\eta_2\nonumber\\
&\hspace{0.5cm}\times\left[\tilde \mu Q^\theta_{\eta\eta_1,\sigma}(\mu)-\tilde \mu Q^\theta_{\eta\eta_2,\sigma}(\mu)
+\sigma \tilde Q^\phi_{\eta\eta_1,\sigma}(\mu)-\sigma \tilde Q^\phi_{\eta\eta_2,\sigma}(\mu)\right]\nonumber\\
&\hspace{0.5cm}\times\left[\tilde \nu Q^\theta_{-\eta\eta_1,\sigma}(\nu)-\tilde \nu Q^\theta_{-\eta\eta_2,\sigma}(\nu)
+\sigma \tilde Q^\phi_{-\eta\eta_1,\sigma}(\nu)-\sigma \tilde Q^\phi_{-\eta\eta_2,\sigma}(r'l)\right]2P_{\eta_1\eta_2}(-\varepsilon\omega_0)\nonumber\\
&=\frac{e^2}{h}\left(\frac{|\Gamma|}{\hbar v}\right)^2
\left(1+\alpha\beta\cos\varphi_u\cos\varphi_l\right)
P(\omega_0)=\tilde S^{(0)}_{\alpha\beta}\,.
\end{align}
In the last line, we used $P(\omega_0)=P_{+-}(\omega_0)+P_{-+}(\omega_0)$. All the temperature and bias dependence is in $P(\omega_0)$.
Therefore,
\begin{equation}
\tilde S^{(0)}_{\alpha\beta}=e\bar I\left(1+\alpha\beta\cos\varphi_u\cos\varphi_l\right)P(\omega_0)\,.
\end{equation}
The correlations between $I^u_{\pm,+(-)}$ and $I^l_{\pm,i}$ vanish. Finally, the correlation between $I^u_i$ and $I^l_i$ is given by
\begin{align}
S^{(i)}_{\alpha\beta}(t,t')
=&-\frac{1}{2\hbar^2}\sum_{\eta}\sum_{\substack{\sigma_1,\varepsilon_1,\eta_1\\ \sigma_2,\varepsilon_2,\eta_2}}
\Gamma^{\varepsilon_1}\Gamma^{\varepsilon_2}\varepsilon_1\varepsilon_2 \sigma_1\sigma_2
\eta_1\eta_2\int_{-\infty}^\infty dt_1\int_{-\infty}^\infty dt_2e^{-i\omega_0(\varepsilon_1t_1+\varepsilon_2t_2)}\nonumber\\
&\times\langle T_c I_0^{u}(l,t,\eta)I_0^{l}(l,t',-\eta)
\psi^{\varepsilon_1}_{u,\sigma_1}(0,t_1,\eta_1)\psi^{\varepsilon_1}_{l,\sigma_1}(0,t_1,\eta_1) \psi^{\varepsilon_2}_{u,\sigma_2}(0,t_2,\eta_2)\psi^{\varepsilon_2}_{l,\sigma_2}(0,t_2,\eta_2)\rangle\nonumber\\
=&\alpha\beta\frac{e^2v_F^2|\Gamma|^2}{8\hbar^2}
\sin\varphi_u\sin\varphi_l
\sum_{\eta,\eta_1,\eta_2,\varepsilon,r=\pm}
\eta_1\eta_2\int_{-\infty}^\infty dt_1\int_{-\infty}^\infty dt_2e^{-i\varepsilon\omega_0(t_1-t_2)}\nonumber\\
&\times\langle T_c \psi_{u,r}^\dagger(rl,t,\eta)\psi_{u,-r}(-rl,t,\eta)
\psi^{\varepsilon}_{u,\varepsilon r}(0,t_1,\eta_1)\psi^{-\varepsilon}_{u,-\varepsilon r}(0,t_2,\eta_2)\rangle\nonumber\\
&\times\langle T_c\psi_{l,r}^\dagger(rl,t',-\eta)\psi_{l,-r}(-rl,t',-\eta)
\psi^{\varepsilon}_{l,\varepsilon r}(0,t_1,\eta_1) \psi^{-\varepsilon}_{l,-\varepsilon r}(0,t_2,\eta_2)\rangle\,.
\end{align}
$l$ is taken to be a reference point located between the end of the interacting region ($x=L$) and the location of the beam splitter. After defining
\begin{align}
C(t-t')=&\frac{|\Gamma|^2ev_F^2}{4\hbar^2\bar I}
\sum_{\eta,\eta_1,\eta_2,\varepsilon,r=\pm}
\eta_1\eta_2\int_{-\infty}^\infty dt_1\int_{-\infty}^\infty dt_2e^{-i\varepsilon\omega_0(t_1-t_2)}\nonumber\\
&\times\langle T_c \psi_{u,r}^\dagger(rl,t,\eta)\psi_{u,-r}(-rl,t,\eta)
\psi^{\varepsilon}_{u,\varepsilon r}(0,t_1,\eta_1)\psi^{-\varepsilon}_{u,-\varepsilon r}(0,t_2,\eta_2)\rangle\nonumber\\
&\times\langle T_c\psi_{l,r}^\dagger(rl,t',-\eta)\psi_{l,-r}(-rl,t',-\eta)
\psi^{\varepsilon}_{l,\varepsilon r}(0,t_1,\eta_1) \psi^{-\varepsilon}_{l,-\varepsilon r}(0,t_2,\eta_2)\rangle
\label{C}
\end{align}
and Fourier transforming the noise, we finally get
\begin{equation}
\tilde S^{(i)}_{\alpha\beta}=\alpha\beta C(\omega_0)e\bar I\sin\varphi_u\sin\varphi_l\,.
\end{equation}
\end{widetext}

%

\end{document}